\documentclass[pra,aps,twocolumn,showpacs,10pt,floatfix]{revtex4-1}
\usepackage{amsmath,amssymb,graphicx,bm,color,ulem}
\usepackage{amsfonts,amsthm}

\begin{document}

\title{Enhanced quantum tunneling in quantum Zeno dynamics freezing momentum direction}
%Enhanced quantum tunneling through a potential barrier by Zeno-freezing the momentum direction}

\author{Miguel A. Porras}
\affiliation{Grupo de Sistemas Complejos, ETSIME, Universidad Polit\'ecnica de Madrid, Rios Rosas 21, 28003 Madrid, Spain}

\author{Nilo Mata}

\affiliation{Departamento de Energ\'{\i}a y Combustibles, ETSIME, Universidad Polit\'ecnica de Madrid, Rios Rosas 21, 28003 Madrid, Spain}

\author{Isabel Gonzalo}

\affiliation{Departamento de \'Optica, Facultad de Ciencias F\'{\i}sicas, Universidad Complutense de Madrid,
Ciudad Universitaria s/n, 28040 Madrid, Spain}

\begin{abstract}
Quantum tunneling is a fundamental quantum mechanical effect involved in plenty of physical phenomena. Its control would impact these phenomena and the technologies based on them. We show that the quantum tunneling probability through a potential barrier can be increased to approach unity in a quantum Zeno dynamics undergone by the tunneling particle in which the direction of the momentum is frequently monitored. We first model the measurements of the momentum direction as selective von Neumann projections, and then as nonselective, direction-sensitive interactions of the particle with probe particles. Nonselective measurements are more efficient than selective measurements in enhancing the quantum tunneling probability.
\end{abstract}

%\pacs{32.80.Wr; 42.65.Tg; 05.45.Yv}

\maketitle

\section{Introduction}

To explain the quantum Zeno effect (QZE) it is often said that ``a watched pot never boils". Frequent measurements that ascertain whether a quantum system remains in its initial state slow down its evolution. The QZE has been known for almost a century \cite{NEUMANN}, was given its name after \cite{MISRA}, and has been observed in diverse physical systems \cite{ITANO,FISCHER}. The simplest manifestations take place in quantum two-level systems; for example, the inhibition of decay of an unstable state \cite{ITANO}, or inhibition of quantum tunneling (QT) in a double-well potential \cite{ALTENMULLER,LERNER}. If the repeated measurements are faster than the short-term parabolic transition probability characteristic of Schr\"odinger evolution, decay or QT become less probable, and impossible with a continuous of measurements. Still, the QZE or ``paradox" is not exempt of controversies, also in the double-well with different models of measurement \cite{PASCAZIO,FEARN,HOME,RUSECKAS,GODBEER}.

A generalization of the QZE is quantum Zeno dynamics (QZD) \cite{FACCHI2}. The frequent measurements of an observable of the system ascertain here whether the state of the system is in a multidimensional subspace of the system's Hilbert space \cite{FACCHI1}. In QZD, the remaining or concomitant QZE (in the sense of freezing the evolution) is more subtle: what tends to be frozen is the transition from the multidimensional subspace defined by the measurements to the complementary subspace, but the system continues to evolve coherently within each subspace \cite{FACCHI3}. QZD has recently been observed in a rubidium condensate, where a superselection rule between two- and three-dimensional subspaces arises \cite{SCHAFER}.

QZD becomes more involved in an infinite-dimensional Hilbert space \cite{FACCHI1}, as that of a simple moving particle. In all previous analysis  the measurements involve position measurements, which, being necessarily imprecise, are aimed to ascertain the location in a region of space, say $\Delta x$. Ref. \cite{FACCHI4} provides a theoretical demonstration that the QZD in the limit of continuous, selective measurements of position, modeled as von Neumann projections, becomes a unitary evolution confined in the Hilbert subspace defined by $\Delta x$ with hard, Dirichlet boundary conditions. With a discrete number $N$ of measurements, a free particle ---the Zeno arrow--- is seen to tend to stop in $\Delta x$ when $N\rightarrow\infty$, forming a quantum-cat-state of the particle moving forward and backward at the same time \cite{PORRAS}. In \cite{WALLACE}, computer simulations of QZD (although it is called QZE) with measurements of position, modeled either as nonselective projections or pointer measurements, are also be seen to freeze the particle. The dynamics of a particle subjected to position measurements and undergoing reflection, localization or exclusion from the measurement region has been studied in detail in Refs. \cite{MACKRORY,GORDON}.

To our knowledge, the complementary QZD with momentum measurements in place of position measurements has not been addressed. Except for the harmonic oscillator, there is no a space-momentum symmetry in the dynamics of a particle, so that QZD with position and with momentum measurements are different dynamical problems. For a free particle, a QZD with projective measurements of position is of interest because each measurement changes the position probability density at the time of the next measurement, but a QZD with measurements of momentum is trivial because the measurements do not change the momentum probability density. Expressed in the terms of modern quantum information and technologies \cite{BRAGINSKY,RALPH,SHOMRONI}, momentum is a quantum nondemolition variable for a free particle whose measurement back action effects can be evaded, and position is not. For a particle subjected to forces, however, momentum is no longer a quantum nondemolition variable, which triggers the rich QZD studied here.

In this paper we show that repeated measurements that would ascertain whether the momentum is in a region of momentum space tend to freeze the momentum in that region. We focus on the effect of this QZD on QT through a potential barrier. To the purpose of tunneling the barrier, it turns out that measurements that simply determine the direction of momentum (positive or negative in our one-dimensional geometry) suffice for the direction to freeze, and therefore for the particle to pass the barrier (see an animation in the supplementary material \cite{SUPPLEMENTAL}).

We first adopt an operational approach to measurements as selective von Neumann projections. For a particle launched towards positive $x$, only the possibility that the momentum is positive in all measurements is considered. Its simplicity allows us to simulate the effect of high enough number of measurements with which the QT probability approaches unity. Next we move to a more physical model of measurement as a nonselective interaction with a probe particle to which information on the direction of the particle is transferred, and reduces the state of the particle to a mixture of states with opposite momenta \cite{SIMONIUS,ZUREK,MENSKY}. In this picture, all possibilities that the momentum is positive at the last measurement are considered. The measurements as interactions also cause the direction to freeze, and the particle to tunnel, with a probability higher than with the selective measurements, and therefore also approaching unity as their frequency increases, pointing to the emergence of a superselection rule between Hilbert's subspaces of different momenta.

Contrary to QZE-induced QT inhibition in a double-well by measuring well occupation, QZD with momentum measurements promotes QT. As such, relevant related phenomena would be those where enhancement of QT will have an impact, such as facilitating nucleus-nucleus collisions in colliders \cite{SALGADO}, or controlling of reaction kinetics \cite{SCHREINER}. In optics, promoting electron tunneling can have an impact in photoelectron ionization generating high harmonics. Vice versa, high harmonics are being employed to probe with attosecond resolution proton dynamics \cite{PROTON} and electrons tunneling a barrier \cite{INTERFEROMETRY,CLOCK}. These breakthroughs would bring closer to hand an implementation of this QZD.

\section{Preliminaries}

We consider a particle of mass $m$, initial state $|\psi_0\rangle$ of wave function $\langle x|\psi_0\rangle =\psi_0(x)=\psi(x)\exp(ik_0 x)$ well-within the half-space $x<0$. The mean momentum $p_0=\hbar k_0$ is positive and such that its kinetic energy $E_c= p_0^2/2m$ is smaller than the peak potential energy $V_0$ of a barrier $V(x)$ localized about $x=0$. For concreteness we will use $V(x)=V_0/(1+ |x/b|^\alpha)$, where $b$ measures the half-width of the barrier, and $\alpha>0$ ($\alpha\rightarrow\infty$ is a square barrier). Of course the details of the tunneling dynamics for different barrier shapes, but the results concerning QT probability under QZD are the same. For a well-directed particle translation dominates over wave packet spreading, implying that $p_0$ is larger than its uncertainty, $\Delta p$.  We rewrite Schr\"odinger equation,
\begin{equation}\label{SCH}
i\hbar \frac{\partial\psi}{\partial t} = -\frac{\hbar^2}{2m}\frac{\partial^2\psi}{\partial x^2} +V\psi\,,
\end{equation}
using the dimensionless spatial coordinate $\xi=x/(\hbar/\Delta p)$ ($x$ in units of the wave function width $\Delta x\sim \hbar/\Delta p$) and dimensionless time $\tau = t/(2m\Delta x /p_0) =t/(2\hbar m/\Delta p p_0)$ (time in units of the time taken for the particle to traverse its own wave function), as
\begin{equation}\label{SCH2}
i\frac{\partial\psi}{\partial \tau} = -\frac{1}{\kappa_0}\frac{\partial^2\psi}{\partial \xi^2} + \kappa_0 v_0 v\psi\,,
\end{equation}
where $v= 1/(1+|\xi/\xi_b|^\alpha)$, and $\xi_b$ is the potential width in units of the wave function width. The parameters at work in the QT dynamics are then $v_0=V_0/E_c$ and $\kappa_0= p_0/\Delta p$. The situation of interest is that with $v_0>1$ (QT) and $\kappa_0>1$ (well-directed particle).

\section{Quantum Zeno dynamics of a tunneling particle with selective measurements of momentum direction}

In contrast to previous research, the QZD considered here is that experienced by the particle subjected to frequent measurements of momentum, i.e., measurements that ascertain whether the momentum is in a certain interval of momentum space or not. However, in the context of QT, the QZD turns out to be qualitatively similar if this interval is $\Delta p$, any other interval about $p_0$, or is simply positive.

In a first approach, the measurements are operationally summarized as projections of the state onto the subspace of positive momenta. These measurements are selective, i.e., meaning that we only consider the event that the momentum is positive in all the measurements. This would be the situation with a measuring device that eliminates the particle upon a negative outcome. In the standard Zeno scheme, the particle evolves according to (\ref{SCH2}) during a certain total time $\tau_{\rm max}$, while the evolution is $N$ times interrupted to monitor the direction of movement. The projective measurements are performed at the equispaced time intervals $\Delta\tau=\tau_{\rm max}/N$. Formally, the state of the particle at the time $\tau_{\rm max}$ is given by
\begin{equation}\label{NORM}
|\phi_N\rangle =\frac{1}{\sqrt{P_N}} \Pi e^{-iH\Delta \tau}\!\!\dots \frac{1}{\sqrt{P_2}} \Pi e^{-iH\Delta \tau} \frac{1}{\sqrt{P_1}} \Pi e^{-iH\Delta \tau} |\psi_0\rangle\,,
\end{equation}
where $H=(-1/\kappa_0)\partial^2\psi/\partial\xi^2 + \kappa_0v_0v(\xi)$ is the Hamiltonian, $\Pi =\int_0^\infty |\kappa\rangle\langle\kappa | d\kappa$ is the projector onto the subspace of positive normalized momenta $\kappa=p/\Delta p$, and $P_n$ is the probability that the momentum is positive at the $n$ intermediate measurement, so that the factors $1/\sqrt{P_n}$ normalize the state after each projective measurement. After the $N$ measurement at time $\tau_{\rm max}$, the probability that all outcomes were positive is the product $P_N^{(s)} = P_1 P_2 \dots P_N$.

To obtain $P_N^{(s)}$, however, it is simpler and faster computationally to evaluate
\begin{equation}\label{UNNORM}
|\psi_N\rangle =\Pi e^{-iH\Delta \tau}\stackrel{^{\mbox{\small $N$ times}}}{\dots\dots} \Pi e^{-iH\Delta \tau} \Pi e^{-iH\Delta \tau} |\psi_0\rangle\,,
\end{equation}
which is obviously related to the normalized state $|\phi_N\rangle$ in (\ref{NORM}) by $|\psi_N\rangle =\sqrt{P_1P_2\dots P_N}\,|\phi_N\rangle$, and whose norm, $\langle\psi_N|\psi_N\rangle =P_1P_2\dots P_N=P_N^{(s)}$, yields directly the probability that all outcomes were positive. If desired, the normalized final state can be calculated from $|\phi_N\rangle =|\psi_N\rangle /\sqrt{P_N^{(s)}}$.

On a computer, each $\Pi e^{-iH\Delta \tau}$ step is performed by solving (\ref{SCH2}) for the wave function in space representation using a split-step fast Fourier transform algorithm, and then eliminating negative momenta by multiplying the wave function in momentum representation by the Heaviside step function. If $\langle \kappa|\psi_N\rangle=\hat\psi_N(\kappa)$ is the momentum wave function at $\tau_{\rm max}$, the probability that all outcomes were positive can be evaluated as
\begin{equation}\label{PNS}
P_N^{(s)}=\langle\psi_N|\psi_N\rangle=\int_{-\infty}^\infty |\hat \psi_N(\kappa)|^2 d\kappa
\end{equation}
(the lower limit can be set to zero due to the last projection at $\tau_{\rm max}$).

To analyze QT under this QZD, we choose $\psi_0(x)$ such that the barrier does not yet affect the particle, and the time $\tau_{\rm max}$ such that the transmission/reflection process is completed (with or without measurements). For a particle launched from the left, the probability of finding the particle to the right of the barrier at $\tau_{\rm max}$, $P_{x>0}^{(s)}= \int_0^\infty |\psi_N(\xi)|^2 d\xi$, must coincide with the probability $P_N^{(s)}$ that the momentum is positive, since the reflected wave function at $x<0$ has only negative momenta. With a single measurement, $P_1^{(s)}$ yields the standard QT probability.

\begin{figure}
  \centering
  \includegraphics*[height=4cm]{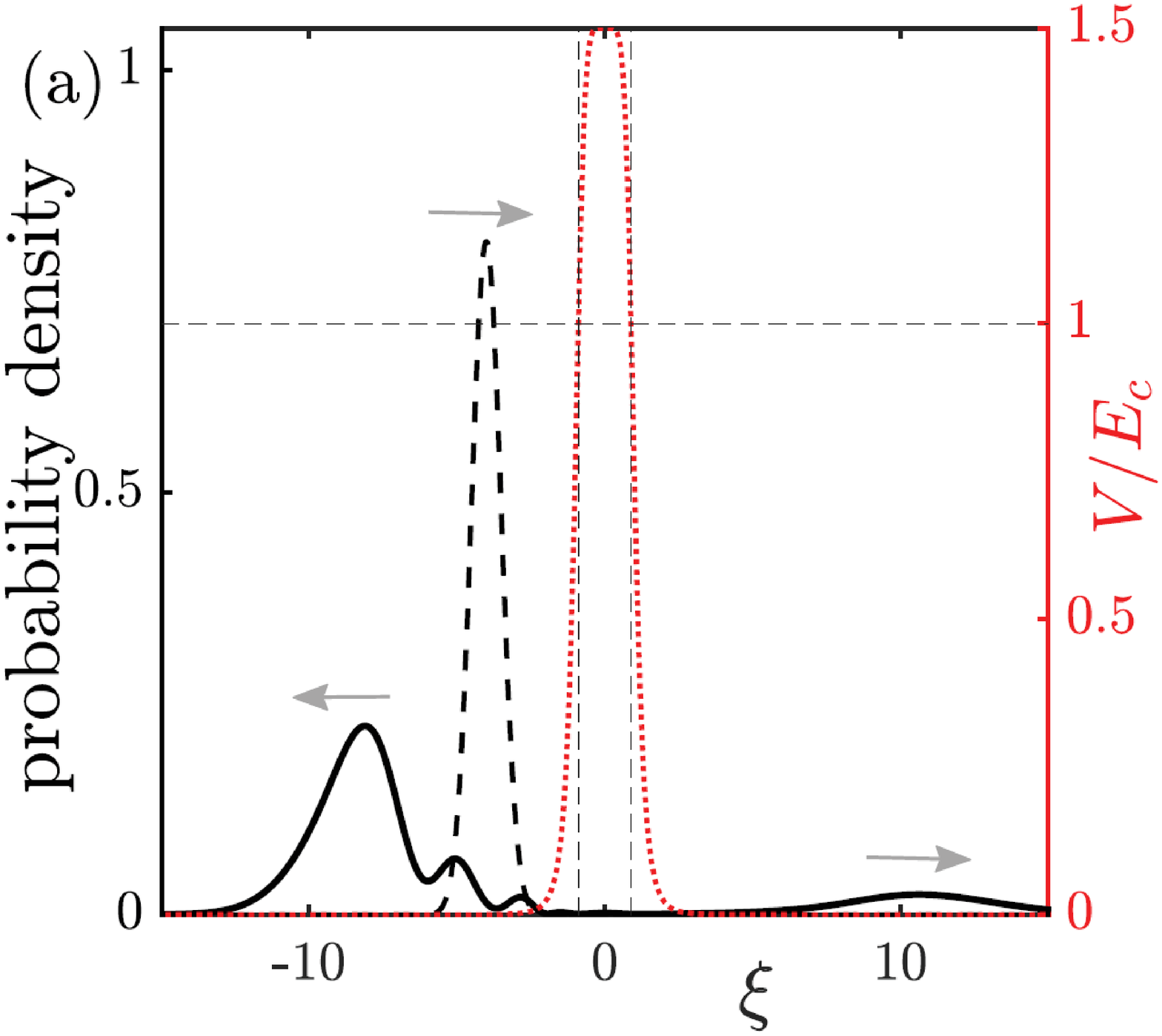}\includegraphics*[height=4cm]{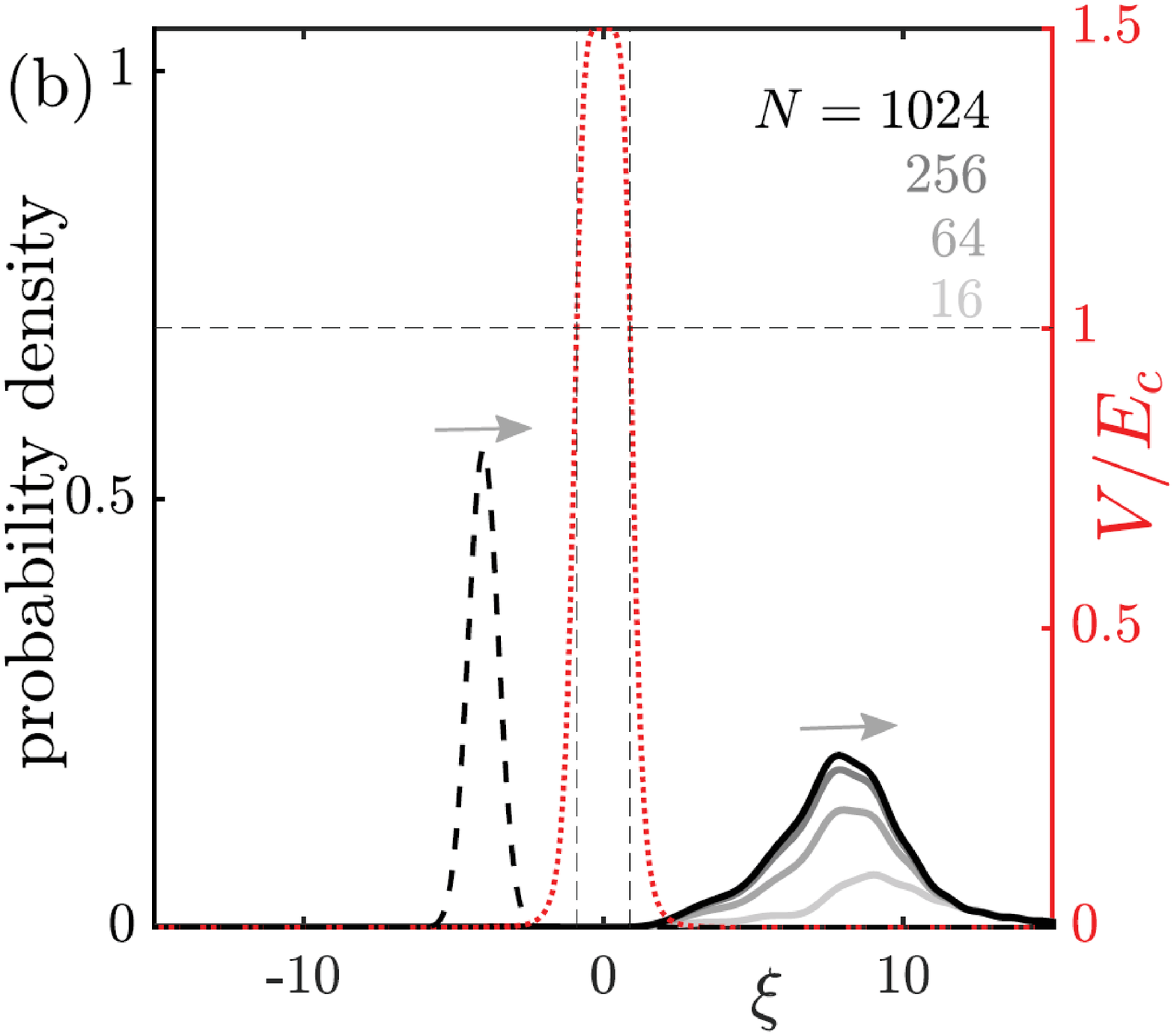}
  \includegraphics*[height=4cm]{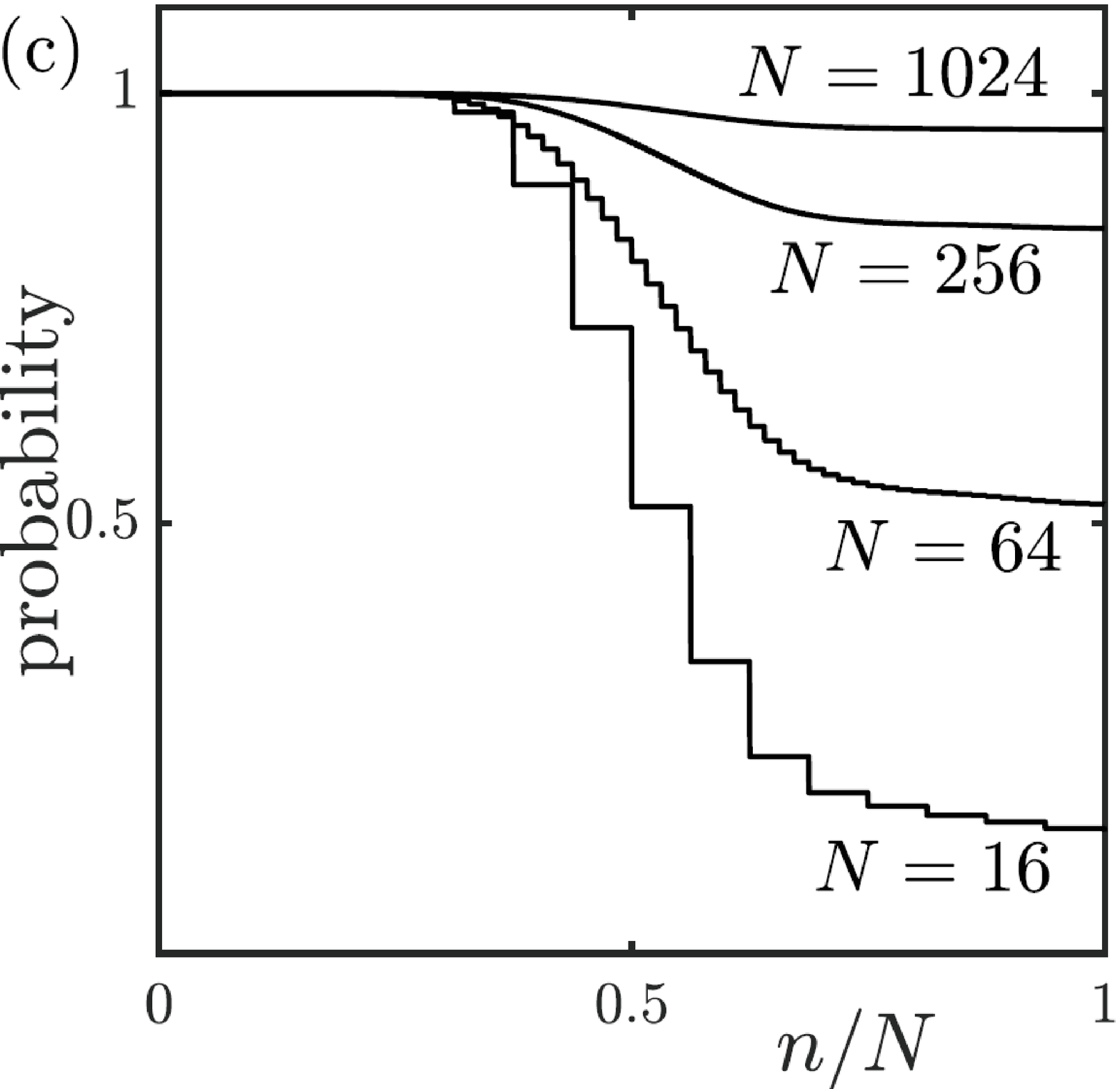}\hspace{0.2cm}\includegraphics*[height=4cm]{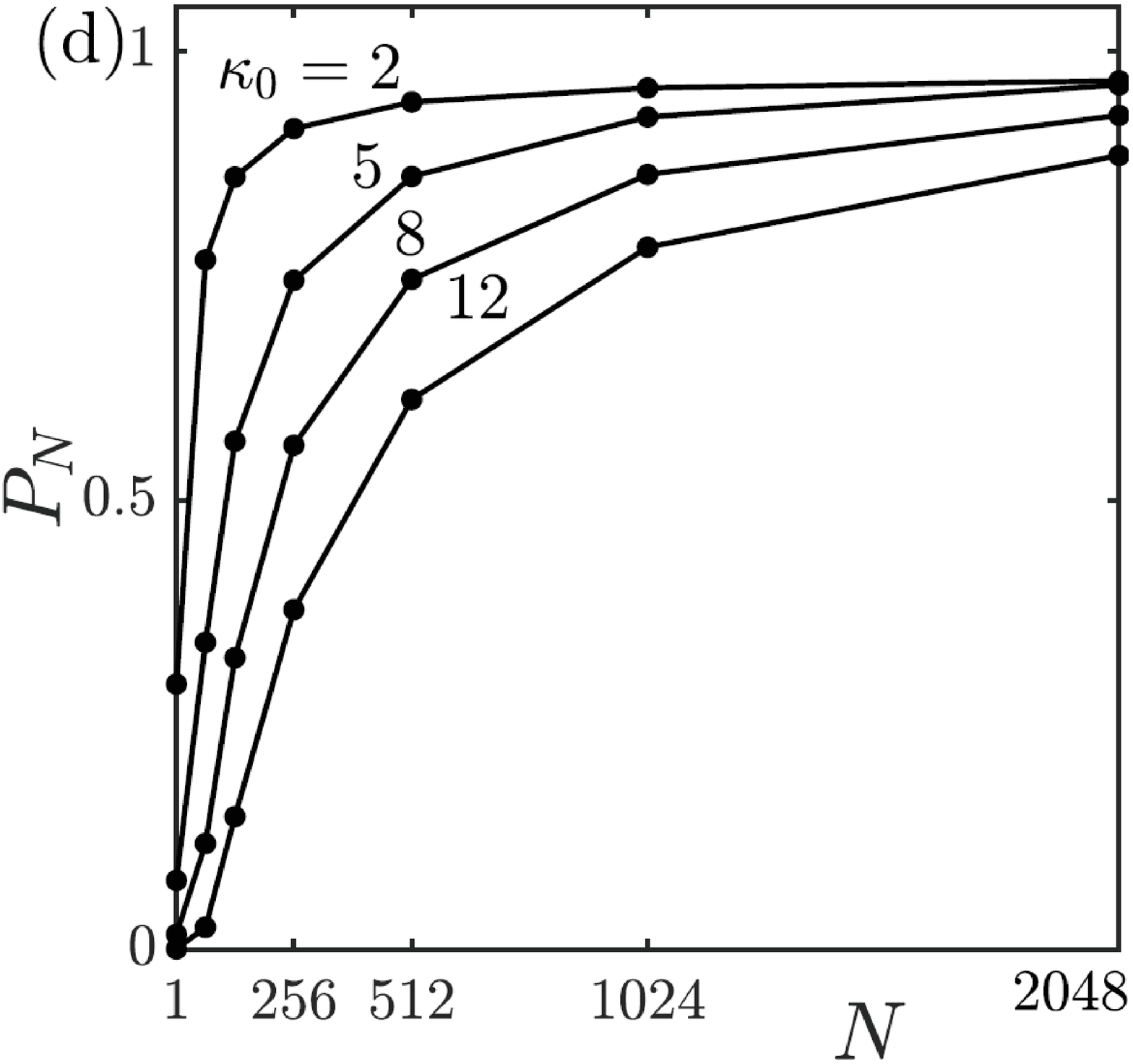}\\
  \caption{\label{Fig1} (a) Partially transmitted and reflected probability density (solid curves) at time $\tau_{\rm max}=6.24$ for the input wave packet $\psi=(2/\pi)^{1/4}\exp{[-(\xi-\xi_0)^2]}\exp(i\kappa_0\xi)$ ($\kappa_0=4$, $\xi_0=-4$) at $\tau=0$ (dashed curve) through the barrier of width $\xi_b=1$, $\alpha=6$ and height $v_0=1.5$ (red dotted curve) without any measurement. Tunneling probability is $P^{(s)}_1=0.162$. (b) The same but with $N=2^4, 2^6, 2^8$ and $2^{10}$ measurements between $\tau=0$ and $\tau_{\rm max}$ yielding respective tunneling probabilities  $P_N^{(s)}=0.31,0.69,0.90$, and $0.97$. The weak reflected wave is eliminated by the projections. The vertical lines indicate the turning points, the horizontal line the energy, and the arrows the movement direction. See animations of (a) and (b) from $\tau=0$ to $\tau_{\rm max}$ in \cite{SUPPLEMENTAL}. (c) Probability that the momentum direction remains positive at the $n$th measurement, for different values of $N$. (d) Tunneling probability as a function of $N$ for several values of $\kappa_0$.}
\end{figure}

In the example of Fig. \ref{Fig1}(a), a Gaussian wave packet (dashed curve) is launched against the potential barrier (dotted curve). Its shape is neither too sharp (high $\alpha$) nor too delocalized (low $\alpha$) to facilitate numerical calculations (less spatial points and temporal steps are required). Without measurements the QT probability is $P_1=0.162$. Figure \ref{Fig1}(b) evidences that monitoring more and more frequently whether the direction remains positive increases the QT probability by freezing the momentum direction. See the QT dynamics in more detail in the videos in the supplementary material \cite{SUPPLEMENTAL}. To illustrate how the increasing tunneling probability $P_N^{(s)}$ is furnished from one to the next measurement, Fig. \ref{Fig1}(c) represents the decreasing probability $P_1P_2\dots P_n$ that the momentum remains positive as the number $n$ of measurements performed increases up to $N$. The larger $N$ (the more steps down), the higher $P_N^{(s)}$ (the less you go down). The QT probability $P_N^{(s)}$ is represented as a function of $N$ in Fig. \ref{Fig1}(d) for different values of $\kappa_0$. As $\kappa_0\rightarrow \infty$, the QT probability without measurements tends to zero, which, given the fixed value of the barrier height $v_0>1$, can be regarded as the classical limit. The QT probability tends always to unity, but it needs more measurements as $\kappa_0$ increases.

\begin{figure}[!t]
\centering
\includegraphics*[height=3.9cm]{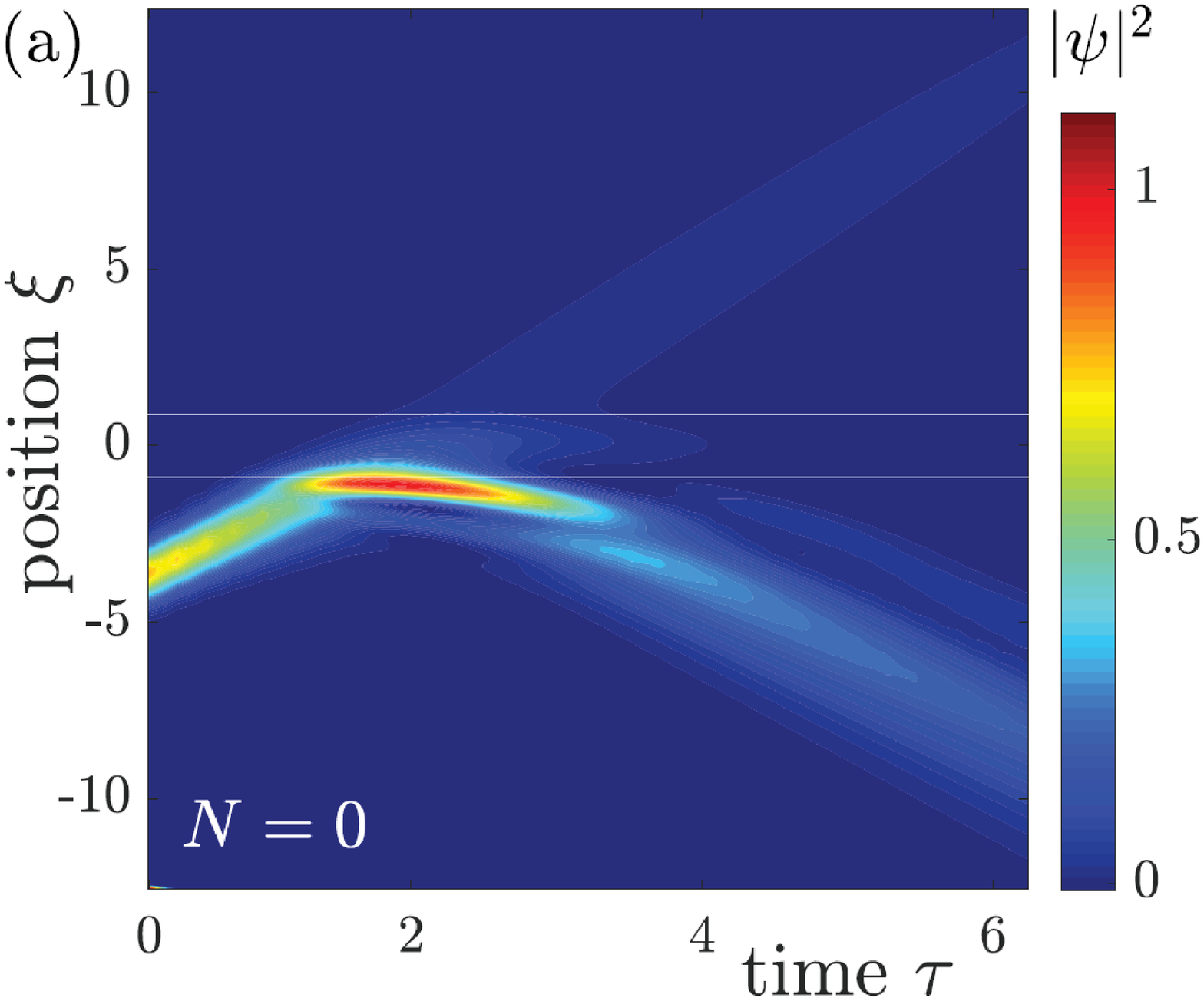}\includegraphics*[height=3.9cm]{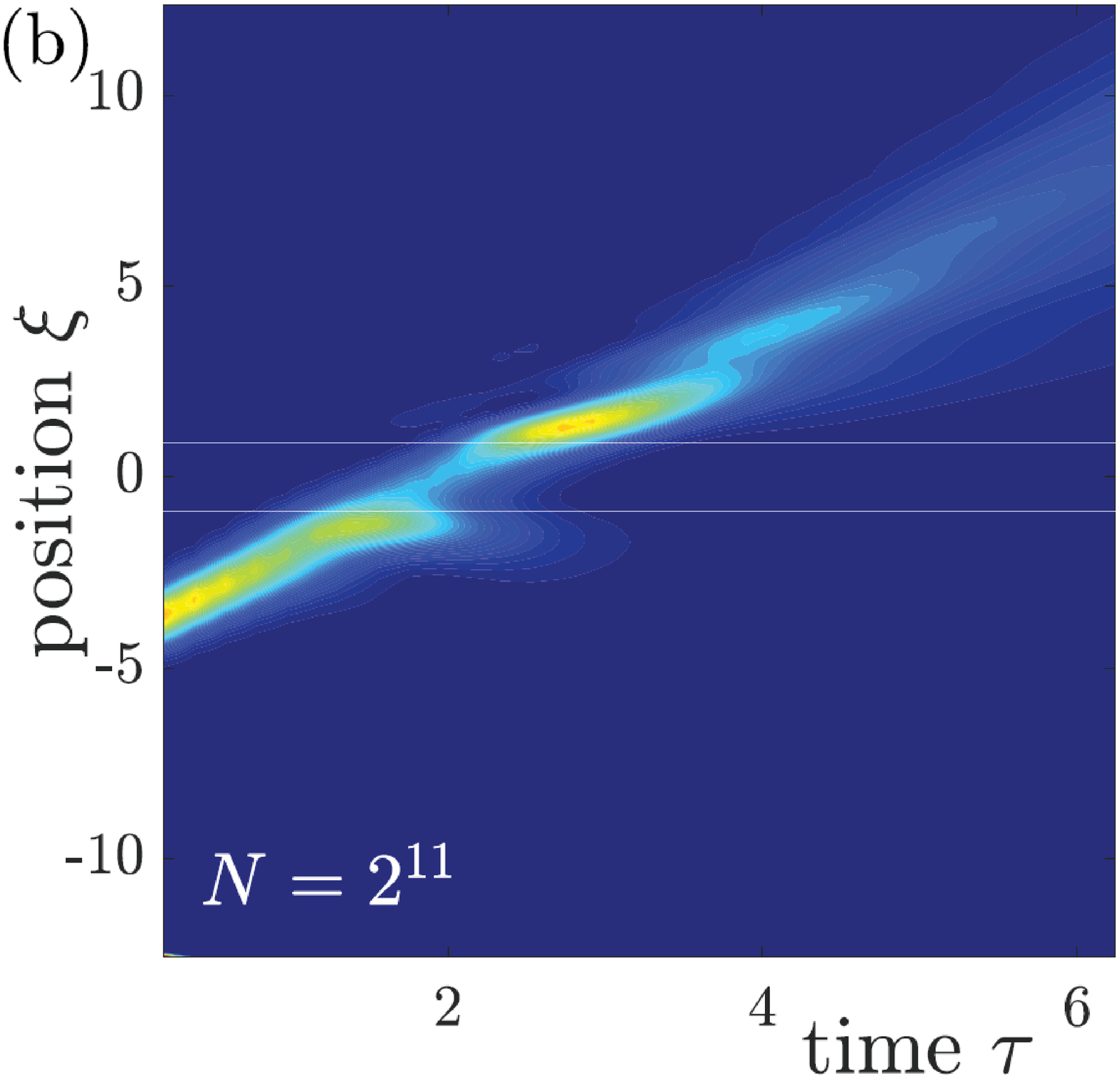}
\caption{\label{Fig2} For the same example as in Fig. \ref{Fig1}, contour plot of the probability density in space and time (a) without measurements and (b) with $N=2^{11}$ measurements.  The horizontal white lines are the classical turning points.}
\end{figure}

The QZD of the tunneling particle, without and with measurements, can be seen unfolded over time in the animations in the supplementary material \cite{SUPPLEMENTAL} and in Fig. \ref{Fig2}. Without measurements (left), the spike that emerges when the wave function reaches the first turning point is mostly reflected and spreads subsequently. With measurements of momentum direction (right), the spike is not reflected, but transforms momentarily into a wide wave packet of low probability density ($\tau\simeq 2.36$) and high positive current under the barrier, and quickly revives in a new spike beyond the second turning point.

\begin{figure}[!b]
\includegraphics*[height=4.0cm]{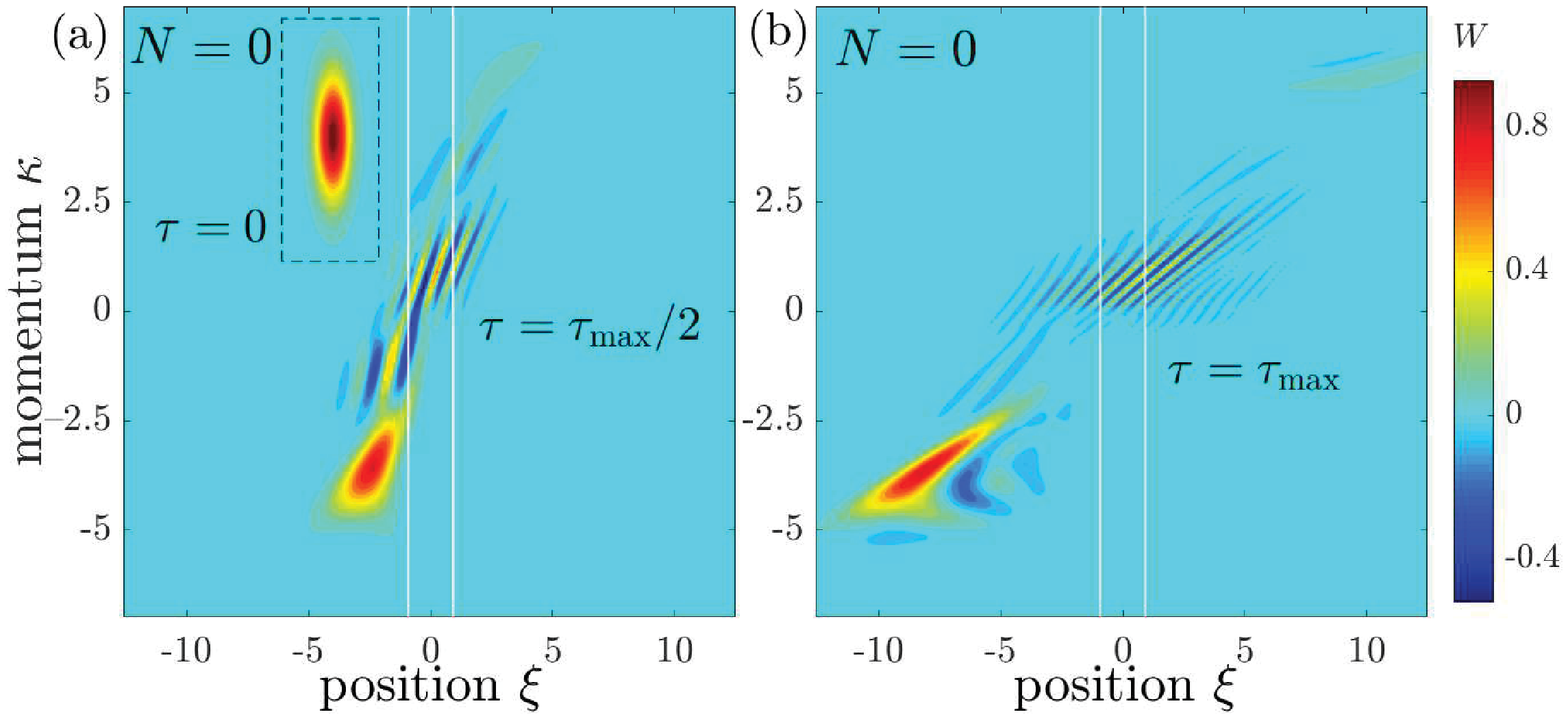}\\ \includegraphics*[height=4.0cm]{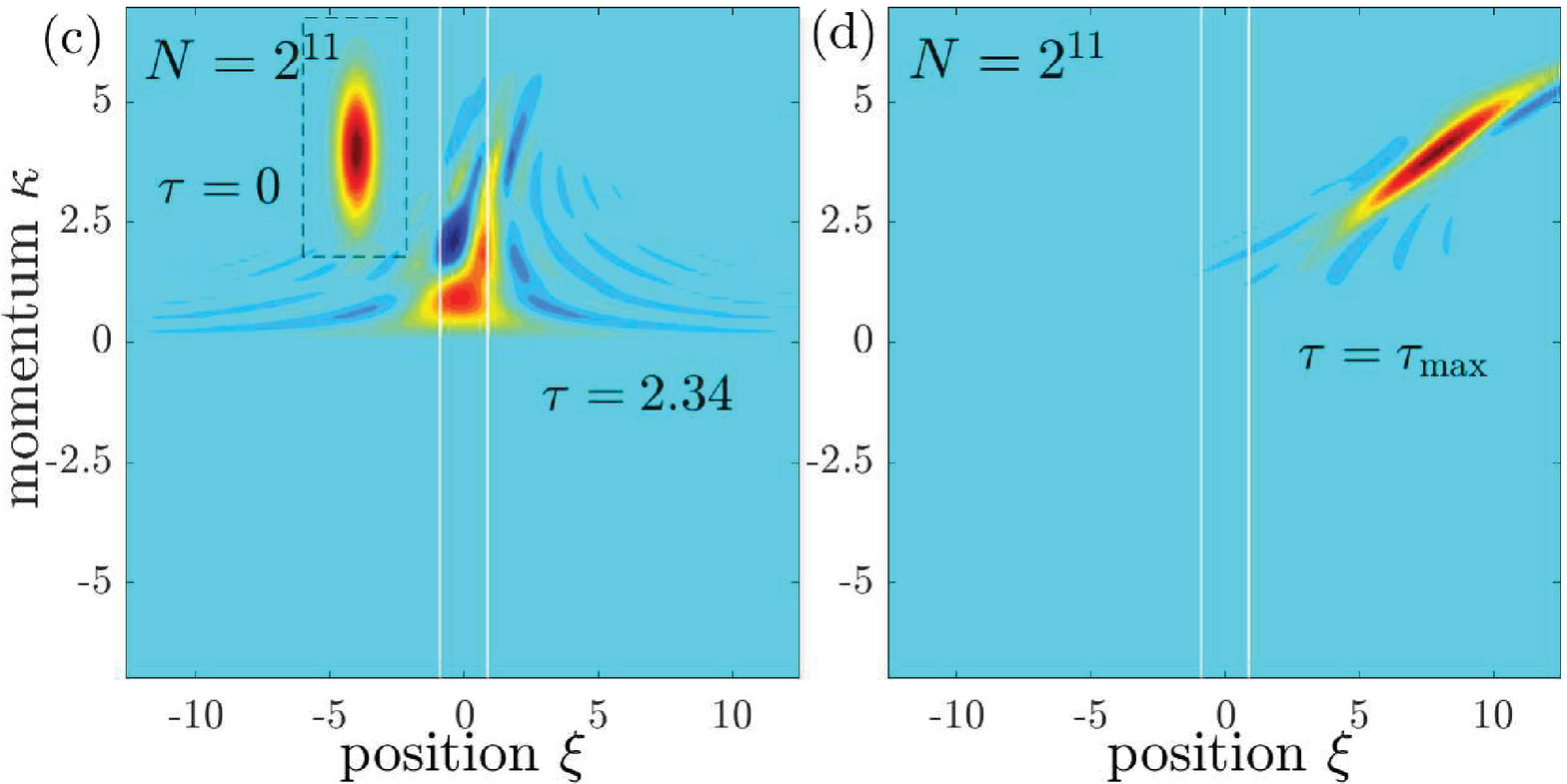}\hspace*{0.65cm}
  \caption{\label{Fig3} For the same example as in Fig. \ref{Fig1}, Wigner distribution function $W(\xi,\kappa)=(1/\pi)\int_{-\infty}^\infty d\eta \psi^\star(\xi+\eta)\psi(\xi-\eta)e^{2i\kappa\eta}$ at the indicated instants of time (a,b) without any measurement, and (c,d) with $N=2^{11}$ measurements. The initial ($\tau=0$) Wigner distribution function is in the box in (a) and (c). The white vertical lines localize the classical turning points. See the evolution in time of the Wigner distribution function in the animation in \cite{SUPPLEMENTAL}.}
\end{figure}

The QZD in phase space can be seen in the animations of the corresponding Wigner distribution functions in the supplemental material \cite{SUPPLEMENTAL}, and in the selected snapshots in Fig. \ref{Fig3}. In contrast to the unmeasured Wigner distribution (top), the frequently measured distribution (bottom) is reflected at zero momentum towards positive momenta  with higher probability as $N$ increases, while the potential barrier substantially ceases to act as such. At tunneling times (e.g. $\tau\simeq 2.36$) the Wigner distribution function of the state of low probability density under the barrier displays not only small ripples but large alternate regions of positive and negative values (blue and red), indicating a highly non-classical behavior.

\section{Quantum Zeno dynamics of a tunneling particle with nonselective interactions measuring momentum direction}

\begin{figure}[t]
\centering
  \includegraphics*[width=8cm]{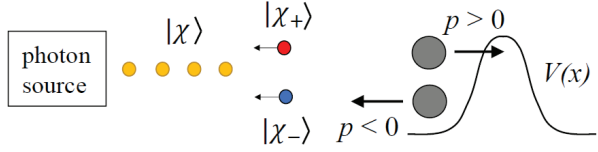}
   \caption{\label{Fig4} Nonselective measurement of momentum direction of an atom by photons. The backscattered photon is red-shifted or blue-shifted depending it finds the atom moving forwards or backwards.}
\end{figure}

These results motivate us to formulate the same problem using a more physical model of the measurement process \cite{SIMONIUS,ZUREK,MENSKY}, which in turn could shed light on how they could be implemented in practice. Each measurement is modelled as a short interaction with an environmental particle, or probe, to which information on the direction of the particle is transferred. Upon tracing over the probe states after the interaction, the positive and negative parts of the wave function decohere.
A possible implementation of these measurements would be the elastic scattering of photons by a moving atom or ion, as sketched in Fig. \ref{Fig4}. Depending on whether a photon is backscattered by the forward or backward moving atom, the photon would be red or blue Doppler shifted, while the atom remains substantially unaltered in its energy and momentum \cite{GUO}. These measurements discriminate between the two directions of movement, but neither direction is selected.

With the same QZD scheme, the initial pure state $|\psi_0\rangle$ evolves to $|\psi_1\rangle = e^{-iH\Delta \tau}|\psi_0\rangle$ in a first time interval $\Delta \tau$. Under the action of the potential barrier and wave packet spreading, $|\psi_1\rangle$ may have acquired positive and negative momenta and hence can always be written as the sum $|\psi_1\rangle = |\psi_{1,+}\rangle + |\psi_{1,-}\rangle$ of the orthogonal vectors
\begin{equation}\label{PSIMASMENOS1}
|\psi_{1,+}\rangle = \int_0^{\infty}\!\!\!\!\!d\kappa \hat\psi_1(\kappa) |\kappa\rangle\,, \quad
|\psi_{1,-}\rangle = \int_{-\infty}^{0}\!\!\!\!\!d\kappa \hat\psi_1(\kappa) |\kappa\rangle\,.
\end{equation}
of only positive or negative momenta. Of course they can be written in terms of normalized states as $|\psi_{1,\pm}\rangle =\sqrt{P_\pm}|\phi_{1,\pm}\rangle$, but as above for selective measurements, the probabilities of positive and negative outcome are stored in $|\psi_{1,\pm}\rangle$ as
\begin{equation*}
\langle \psi_{1,+}|\psi_{1,+}\rangle =P_{+},\quad \langle \psi_{1,-}|\psi_{1,-}\rangle =P_{-},
\end{equation*}
The vectors $|\psi_{1,\pm}\rangle$ are directly accessible in momentum representation as $\hat \psi_{1,\pm}(\kappa) = \hat\psi_1(\kappa)\theta(\pm \kappa)$, where $\theta(\cdot)$ is the Heaviside step function, without computing any integral, which saves computational time.

Interaction with the probe, whose initial state is $|\chi\rangle$, leads to the transitions $|\psi_{1,\pm}\rangle|\chi\rangle \rightarrow |\psi_{1,\pm}\rangle|\chi_\pm\rangle$ of the system. Discrimination between positive and negative momenta of the particle requires the states $|\chi_+\rangle$ and $|\chi_-\rangle$ of the probe to be orthonormal. For the state $|\psi_1\rangle$ at $\Delta\tau$, the result of the interaction is the entangled state $|\psi_{1,+}\rangle|\chi_+\rangle + |\psi_{1,-}\rangle|\chi_-\rangle$, whose density matrix is
\begin{eqnarray}\label{RHO}
  \rho_{\rm 1, \rm meas}&=& \left(|\psi_{1,+}\rangle|\chi_+\rangle + |\psi_{1,-}\rangle|\chi_-\rangle\right) \nonumber \\
  &\times& \left(\langle\psi_{1,+}|\langle\chi_+|+ \langle\psi_{1,-}|\langle\chi_-|\right)\,,
\end{eqnarray}
and, if only the particle is considered after separation of the probe, its reduced density matrix is obtained by tracing over the probe:
\begin{eqnarray}\label{RHORED}
\rho_{1,\rm red} &=&\langle\chi_+|\rho_{\rm meas}|\chi_+\rangle + \langle\chi_-|\rho_{\rm meas}|\chi_-\rangle \nonumber \\
       &=& |\psi_{1,+}\rangle\langle \psi_{1,+}| + |\psi_{1,-}\rangle\langle \psi_{1,-}|\,,
\end{eqnarray}
which is a incoherent mixture of a particle travelling forward or backward with probabilities equal to the respective norms.

The particle is next left to evolve another time $\Delta\tau$ to $\rho_2=e^{-iH\Delta \tau}\rho_{1,\rm red}e^{iH\Delta \tau}$, that is,
\begin{eqnarray}
\rho_2&=&e^{-iH\Delta \tau}|\psi_{1,+}\rangle\langle \psi_{1,+}|e^{iH\Delta \tau} \nonumber \\
&+&
e^{-iH\Delta \tau}|\psi_{1,-}\rangle\langle \psi_{1,-}|e^{iH\Delta \tau}\,,
\end{eqnarray}
where, as before, $e^{-iH\Delta \tau}|\psi_{1,+}\rangle \equiv |\psi_{2,+}\rangle = |\psi_{2,++}\rangle + |\psi_{2,+-}\rangle$ and $e^{-iH\Delta \tau}|\psi_{1,-}\rangle \equiv |\psi_{2,-}\rangle =  |\psi_{2,-+}\rangle +  |\psi_{2,--}\rangle$, since, again, any of these states may acquire opposite momenta. Here, also as above,
\begin{eqnarray}\label{PSIMASMENOS2}
|\psi_{2,++}\rangle &=& \!\!\int_0^{\infty} \!\!\!\!\! d\kappa \hat\psi_{2,+}(\kappa) |\kappa\rangle, \quad
|\psi_{2,+-}\rangle = \!\!\int_{-\infty}^{0} \!\!\!\!\! d\kappa \hat\psi_{2,+}(\kappa) |\kappa\rangle, \nonumber\\
|\psi_{2,-+}\rangle &=& \!\!\int_0^{\infty} \!\!\!\!\! d\kappa \hat\psi_{2,-}(\kappa) |\kappa\rangle, \quad
|\psi_{2,--}\rangle = \!\!\int_{-\infty}^{0} \!\!\!\!\! d\kappa \hat\psi_{2,-}(\kappa) |\kappa\rangle , \nonumber
\end{eqnarray}
whose norms
\begin{eqnarray*}
\langle \psi_{2,++}|\psi_{2,++}\rangle =P_{+}P_{++},\quad \langle \psi_{2,+-}|\psi_{2,+-}\rangle =P_{+}P_{+-}, \\
\langle \psi_{2,-+}|\psi_{2,-+}\rangle =P_{-}P_{-+},\quad \langle \psi_{2,--}|\psi_{2,--}\rangle =P_{-}P_{--},
\end{eqnarray*}
provide the probabilities of realization of the four events. The density matrix is then written as
\begin{eqnarray}
\rho_2 &=& (|\psi_{2,++}\rangle + |\psi_{2,+-}\rangle ) (\langle\psi_{2,++}|+\langle\psi_{2,+-}|)\nonumber \\
 &+& (|\psi_{2,-+}\rangle + |\psi_{2,--}\rangle )  (\langle\psi_{2,-+}|+\langle\psi_{2,--}|) \,,
\end{eqnarray}
and in a second interaction with a probe ($|\psi_{2,\dots \pm}\rangle|\chi\rangle \rightarrow |\psi_{2,\dots \pm}\rangle|\chi_\pm\rangle$) the density matrix of the particle-probe system is
\begin{eqnarray}
\rho_{2,\rm meas} &=& (|\psi_{2,++}|\chi_+\rangle\rangle + |\psi_{2,+-}\rangle|\chi_-\rangle ) \nonumber \\
 & \times & (\langle\psi_{2,++}|\langle\chi_+|+\langle\psi_{2,+-}|\langle\chi_-|)\nonumber \\
 &+& (|\psi_{2,-+}\rangle |\chi_+\rangle+ |\psi_{2,--}\rangle|\chi_-\rangle ) \nonumber \\
 & \times & (\langle\psi_{2,-+}|\langle\chi_+|+\langle\psi_{2,--}|\langle\chi_-|) \,.
\end{eqnarray}
Tracing over the probe after the interaction, the reduced density matrix of the particle is
\begin{eqnarray}
\rho_{2, \rm red}&=&|\psi_{2,++}\rangle\langle \psi_{2,++}| + |\psi_{2,+-}\rangle\langle \psi_{2,+-}| \nonumber \\
      &+& |\psi_{2,-+}\rangle\langle \psi_{2,-+}| +|\psi_{2,--}\rangle\langle \psi_{2,--}|\,,
\end{eqnarray}
which is a incoherent mixture of four states of four possible outcomes in the two measurements with norms equal to the respective probabilities of realization.

\begin{figure}[b]
  \centering
  \includegraphics*[height=4cm]{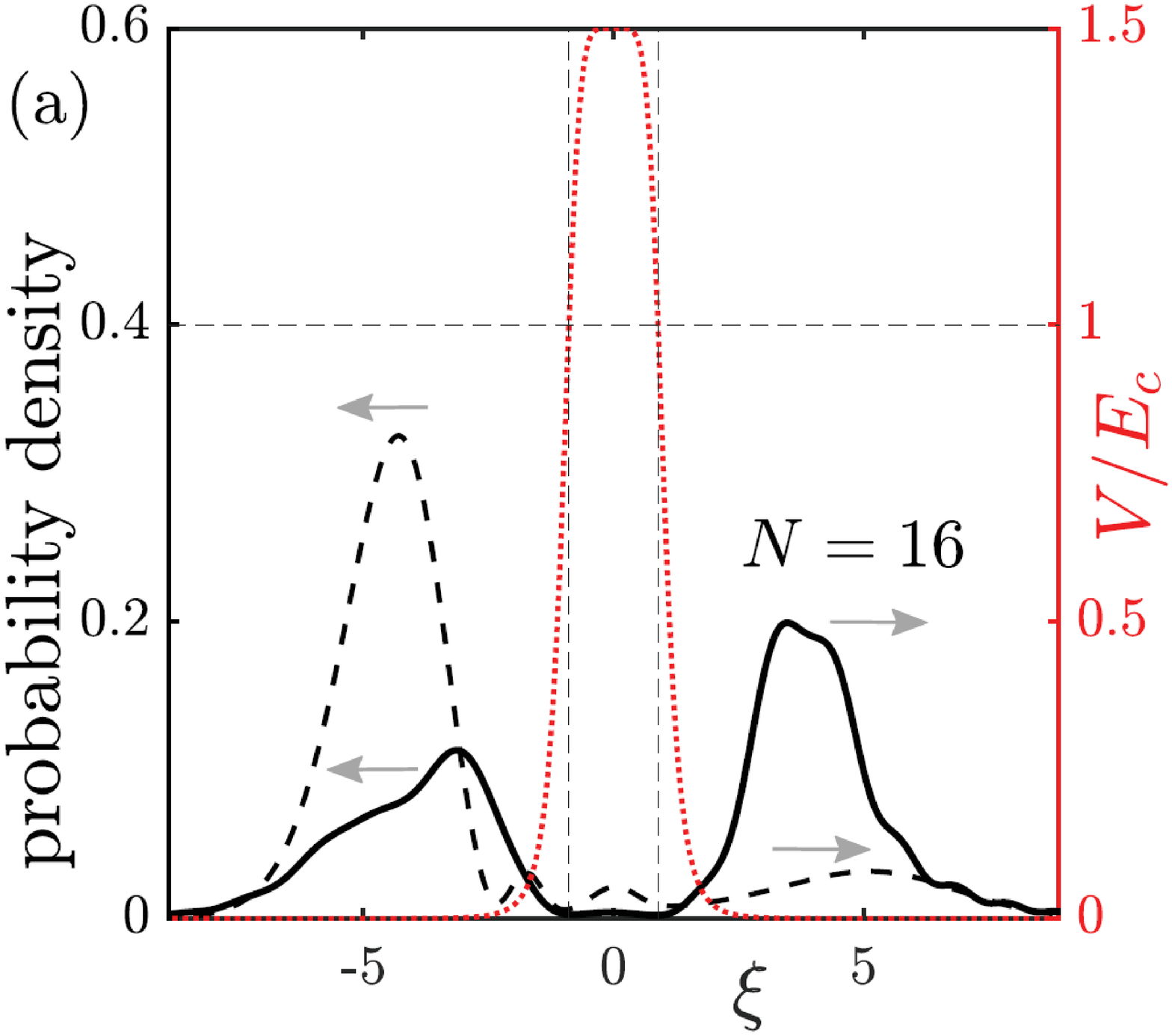}\includegraphics*[height=4cm]{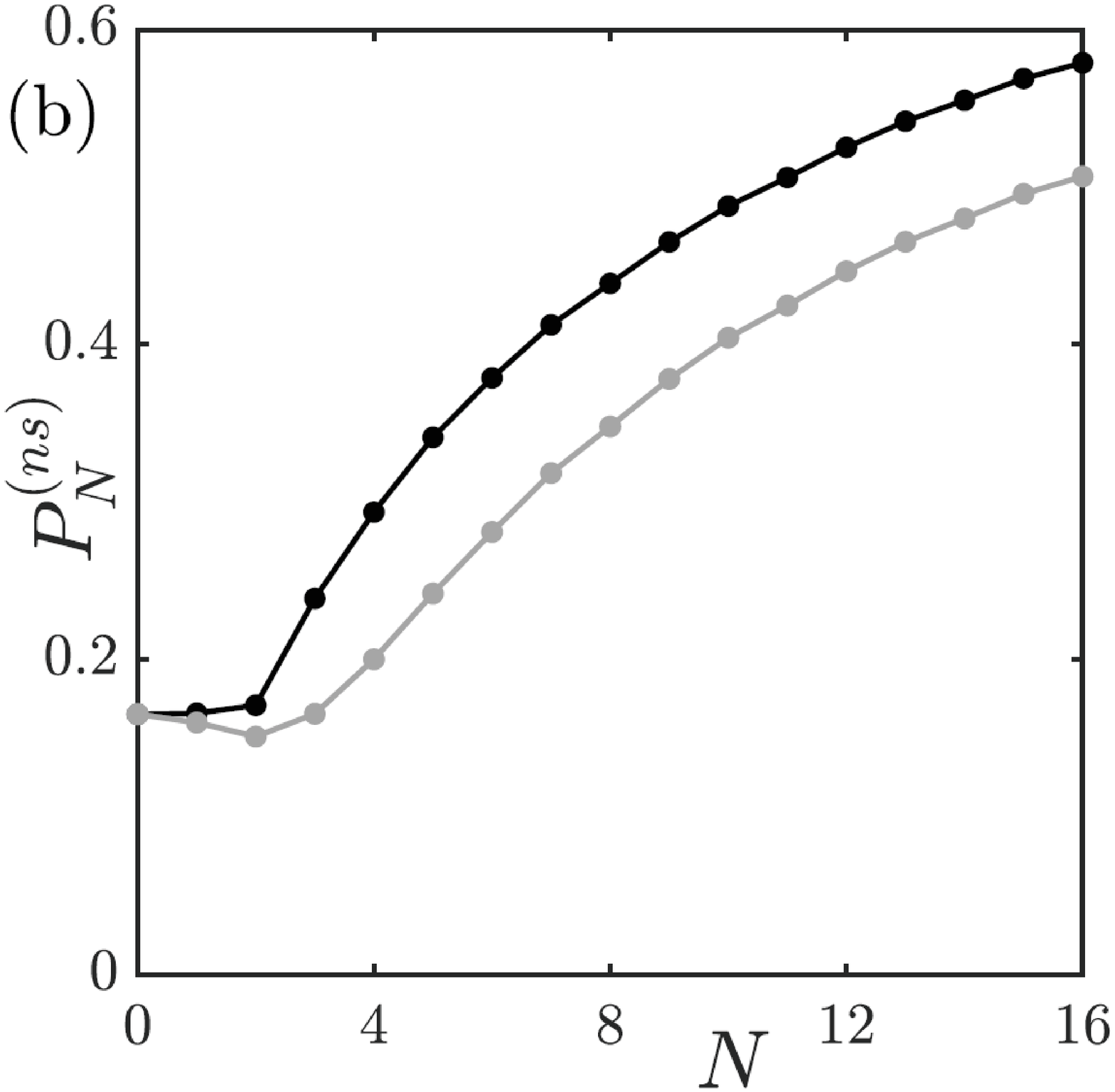}
  \caption{\label{Fig5} (a) Partially transmitted and reflected probability density at time $\tau_{\rm max}=3.56$ for the input wave packet $\psi=(2/\pi)^{1/4}\exp{[-(\xi-\xi_0)^2]}\exp(i\kappa_0\xi)$ ($\kappa_0=4$, $\xi_0=-2.67$) at $\tau=0$ through the barrier of width $\xi_b=1$, $\alpha=6$ and height $v_0=1.5$ (red dotted curve) without any measurement (dashed curve), and the same with $N=16$ measurements (solid curve). To concentrate the measurements, 15 of them are performed between $\tau_1=0.88$ and $\tau_2=3$, and the last one at $\tau_{\rm max}$. QT probability without measurements is $P^{(ns)}_1=0.162$, and with $N=16$ measurements is $P^{(ns)}_1=0.579$. The vertical lines indicate the turning points, the horizontal line the energy, and the arrows the movement direction. (b) QT probability as a function of $N$ when measurements are nonselective (black curve) and selective (gray curve), for comparison.}
\end{figure}

This procedure is repeated up to $N$ times, whereby a mixture of $2^N$ states, e.g.,  $|\psi_{N,++-++\dots +-}\rangle$, (the number of plus and minus signs is $N$) corresponding to the $2^N$ possible outcomes in each of the $N$ measurements is obtained, with probabilities equal to their norms. As with selective measurements, the $2^N$ normalized states, $|\phi_{N,++-++\dots +-}\rangle$, can be obtained by dividing $|\psi_{N,++-++\dots +-}\rangle$ by $\sqrt{\langle\psi_{N,++-++\dots +-}|\psi_{N,++-++\dots +-}\rangle}$, if desired.

Among these states, there are $2^{N-1}$ states yielding positive direction at the last measurement. Thus, the probability that the direction of the particle is positive at the $N$ measurement is the sum of their respective probabilities:
\begin{equation}\label{PN}
P^{(ns)}_N=\sum^{2^{N-1}} \langle\psi_{N,\dots +}|\psi_{N,\dots+}\rangle = \sum^{2^{N-1}}\!\!\int_{-\infty}^\infty \!\!|\hat\psi_{N,\dots+}(\kappa)|^2 d\kappa
\end{equation}
(again the lower limit can be set to zero). Note that the term with $N$ positive signs is the probability that all outcomes were positive and therefore coincides with $P_N^{(s)}$ in (\ref{PNS}). We then conclude that these interactions of the particle with probe particles discriminating the particle direction also result in freezing the direction, and their efficiency is higher than with selective measurements.

If, for brevity we call each incoherent state at the $N$th measurement $|\psi_{N,i}\rangle$, the probability density of finding the particle at $\xi$ is $\langle \xi|\rho_{N,\rm red}|\xi\rangle=\sum_{i}^{2^N}|\langle \xi|\psi_{N,i}\rangle|^2 = \sum_{i}^{2^N} |\psi_{N,i}(\xi)|^2$, verifying $\int_{-\infty}^\infty \langle \xi|\rho_{N,\rm red}|\xi\rangle d\xi=1$ since temporal evolution and measurements are unitary transformations. For a particle initially located well-before the barrier, and with a sufficiently long time $\tau_{\rm max}$, the probability that the particle is to the right of the barrier,
\begin{equation}\label{PROB}
  P_{x>0}^{(ns)} =\int_0^\infty \langle \xi|\rho_{N,\rm red}|\xi\rangle d\xi = \sum_{i}^{2^N}\int_0^\infty |\psi_{N,i}(\xi)|^2 d\xi\,,
\end{equation}
must coincide with $P_N^{(ns)}$, implying also an enhanced QT probability.

We have implemented the above procedure on a computer and relevant results are depicted in Fig. \ref{Fig5}. Note that the number of times Schr\"odinger equation (\ref{SCH2}) is to be solved with $N$ measurements is $2^0$ up to the first measurement, $2^1$ up to the second, and so on, i.e., $\sum_{n=0}^{N} 2^n=2^{N+1}-1$, which grows exponentially and limits the maximum number of measurements to a few tens in our computer facility. Given the difficulty of simulating a high number of measurements, and that they have no effect on the freely moving particle before and after the barrier, the $N$ measurements between $\tau=0$ and $\tau_{\rm max}$ are concentrated in $N-1$ measurements in a shorter time interval $[\tau_1,\tau_2]$ in which the wave function significantly overlaps the potential, plus the $N$ measurement at $\tau_{\rm max}$ for the reflection/transmission process to be concluded. The interval $[\tau_1,\tau_2]$ is the same for all $N$, so that increasing $N$ is the same as increasing the frequency. Under the conditions specified in the caption, Fig.  \ref{Fig5}(a) shows the spatial probability densities with $N=1$, with QT probability $P_1^{(ns)}=0.162$, and with $N=16$ measurements, with QT probability $P_{16}^{(ns)}=0.579$. The plot of $P_N^{(ns)}$ versus $N$ in Fig. \ref{Fig5}(b) evidences the increasing probability that the momentum direction is frozen Zeno effect, and hence the increasing QT probability. Since $P_N^{(s)}$ (gray curve) approaches unity, $P_N^{(ns)}$ (black curve) does faster.

\section{Concluding remarks and conclusion}

Before concluding, we clarify that both QT inhibition in QZD with position measurements (well occupation) \cite{ALTENMULLER,LERNER} and the present QT enhancement in QZD dynamics with momentum measurements involve Zeno effects: they slow down or freeze the magnitude being observed. They tend to generate superselection rules that prohibit transitions between the respective subspaces and their complementary.

On the other hand, anti-Zeno enhancement of QT by position measurements has also been described \cite{AL}, usually associated with less frequent measurements. Analogously, an anti-Zeno inhibition of QT by measurements of momentum direction appears to exist as well, as can be appreciated in the initially decaying gray curve of Fig. \ref{Fig5}(b) for low number of measurements. A detailed investigation of this anti-Zeno effect is deferred to future work.

To conclude, we have unveiled the existence of a quantum mechanical dynamics of a particle ---a QZD involving frequent measurements of momentum direction--- that deeply affects QT up to the point of ensuring transmission. Given the enormous advances in attosecond science (see for example \cite{INTERFEROMETRY,CLOCK} and references therein for all-optical imaging of tunneling electrons), ``Zeno-assisted tunneling" could be implemented in practice, which would impact the diverse phenomena where QT is a key player.

M.A.P. acknowledges support from Project No. PGC2018-093854-B-I00, and M.A.P. and I.G. from project No. FIS2017-87360-P of the Spanish Ministerio de Ciencia, Innovaci\'on y Universidades. N.M. acknowledges support from grant No. D480 (Beca de colaboración de formación) of the Universidad Polit\'ecnica de Madrid.

\end{document}